\documentclass[12pt]{article}

\usepackage{epsfig}
\usepackage{subfigure}
\usepackage{calc}
\usepackage{amssymb}
\usepackage{amstext}
\usepackage{amsmath}
\usepackage{amsthm}
\usepackage{pslatex}
\usepackage[small]{caption}
\usepackage{algorithmic,algorithm}

\newtheorem{definition}{Definition}

\begin{document}

\title{On attribute-based usage control policy ratification for cooperative computing context}
\author{
        Ziyi Su\dag, Frederique Biennier* \\
          Dep. Computer Science, Northeast Normal University\dag\\
          Lab. LIRIS, CNRS, INSA-Lyon*
}

\date{\today}

\maketitle

\begin{abstract}
In an open information systems paradigm, real-time context-awareness is vital for the success of cooperation, therefore dynamic security attributes of partners should considered in coalition for avoiding security conflicts. Furthermore, the cross-boundary asset sharing activities and risks associated to loss of governance call for a continuous regulation of partners' behavior, paying attention to the resource sharing and consuming activities.
This paper describes an attribute-based usage control policy shceme compline to this needs. A concise syntax with EBNF is used to summarize the base policy model. The semantics of negotiation process is disambiguated with abductive constraint logic programming (ACLP) and Event Calculus (EC). Then we propose a policy ratification method based on a policy aggregation algebra that elaborate the request space and policy rule relation. This method ensures that, when policies are aggregated due to resource sharing and merging activities, the resulting policy correctly interprets the original security goals of the providers' policies.
\end{abstract}

\section{Introduction}\label{intro}
Recent years have seen the development of network-based, distributed and cooperative information systems paradigm, such as Web Service, Cloud Computing, Collaborative Enterprise or Social Network settings. The major hallmark of security engineering for such systems is to protect each organization's autonomy and patrimony value while building a consistent multi-organizational business federation. Such an open-system paradigm introduces the needs to control accesses by previous unknown users \cite{Park2004}, probably identified with multiple attributes that reflect their security profiles \cite{Lee2011,Cirio2007}. Besides, the cross-boundary asset sharing activities and risks associated to loss of governance \cite{kagal10Jul} call for a continuous protection covering the full lifecycle of assets, paying attention to the sharing and consuming activities. Such requirements are beyond the expressivity of traditional access control and, therefore, continuous 'due usage'-aware security policy models have been visioned by the research community \cite{Zhang2008a,Park2004,OASIS}. Summarized as Usage Control \cite{Park2004}, such a model extends traditional access control systems (where the decision of policy negotiation process is usually the permission or negation of 'access' action) with necessary elements to express variable privileges upon the object and, especially, obligations \cite{DBLPBCCKKLLLNRW09,Cuppens12} associated to the privileges. It allows providers to define in what manner their assets are consumed and received research attentions since its introduction \cite{Hu2009, Russello2009, Gheorghe2010b}.

%But when applying to cooperative context, modification and adaption are needed on canonical usage control model in order to fit the openness and dynamicity in such context. Firstly, entities in open system paradigm are usually evaluated by multiple attributes when forming business coalition. For example, recent years have seen the rapid development of attribute-based access control perspective \cite{Wang-2004-p45-55,Wang2009a,Zhang2005,Cirio2007}, the representative work being XACML \cite{OASIS}, where entities, i.e. providers, assets and consumers, are described with their attributes.

In an open information systems paradigm, real-time context-awareness is vital for the success of cooperation, therefore dynamic security attributes of partners are considered in coalition. Usage control model has the agility to build authorization condition on identity, role or attributes, thus settles security factors in various system paradigms.

As in cooperative information systems partners' interactions usually incur the aggregation of their assets (we name it as 'O-Asset' in the following discussion), their policies are merged to regulate the further usage of the merged asset (denoted as 'C-Asset' in the following). Therefore a policy ratification process is usually necessary to ensure that partners' security policies are not conflicting. Aware of such issues, this paper discusses the employment of attribute-based usage control policy for compliance-assurance for partner behavior in a distributed cooperative system:

\begin{itemize}
\setlength{\itemsep}{1pt}
\setlength{\parskip}{0pt}
\setlength{\parsep}{0pt}
\item We use a concise syntax with EBNF to summarize the attribute-based usage control policy model.
\item We describe the semantics of negotiation process, with abductive constraint logic programming (ACLP) and Event Calculus (EC) \cite{Kowalski-1986-p67-95}, to formulate the mechanism for matching attribute predicates between requests and rules, as well as the impact of rule/policy combining algorithms. Researchers have described temporal constrains in usage control model, paying particular attention on 'right' exertion \cite{Janicke-2008-p111-118}, 'obligation' \cite{DBLPPretschnerRSW09} and 'attribute updating' \cite{Zhang-2005-p351-387} operations, with Temporal Logics (e.g. TLA). Our discussion, otherwise, focuses on formulating the request-policy matching mechanism, which builds up the foundation for policy ratification mechanism.

\item We propose a policy ratification method based on a policy aggregation algebra to ensure that the resulting policy correctly interprets the original goals of the providers' policies. Very recently an integration algebra \cite{RLBLL2011} has been brought forward to capture the semantics of policy ratification w.r.t. the request space. Since its introduction, it has been employed by researchers to articulate policy analysis processes. Our policy aggregation algebra is also based on this algebra but an extension is made by our own perspective: we describe the logic foundation for policies from different security domains to be correlated. This is based on the awareness that domain-separation and lack of semantic coordination on elements in request spaces hinder the policy conflicts detection and inconsistency resolution process, due to semantic misunderstanding. Our ratification process combines policies with 'specificity precedence' (i.e. rules applying to more specific entity sets take precedence) principle and 'deny override' (deny rules take precedence) combinator. It also allows potential extension for weighted policy combination, simply put: more important polices dominate. After confliction detection, a recommendation method based on 'majority precedence' is used to facilitate choosing more eligible partners from those having positive aggregation results.
\end{itemize}

After discussing the background and related work for the development of attribute-based usage control in section (\ref{context}), we summarize the attribute-based usage control model in section (\ref{Toward}).
The semantics of policy evaluation and aggregation processes are presented in section (\ref{Negotiation}) and (\ref{UCONfC}).

\section{Background and related work}\label{context}
Access control is an approach using explicit rule-based models to grant a right to a subject for the accessing of an object \cite{Park2004}, built around the academic perspective of the access control matrix model \cite{AMitx1971}. Canonical access control models, e.g. mandatory access control (MAC), discretionary access control (DAC) and role-based access control (RBAC), are based on the organizational view and focus on providing authorization for asset accessing in a closed system environment \cite{Park2004}. Such models traditionally use a single credential (identity, role, etc.) to identify a system entity (e.g. subject, object, right, etc.).

\subsection{Usage control}\label{ucon}
The foundation of usage control system has been laid down by some pioneering researches. In the $UCON_{ABC}$ model \cite{Zhang2008a,Park2004}, access decision is based on not only role (as in RBAC) or identity (as usually in Access Control List, MAC and DAC), but on diverse attributes of subjects and objects. With the grant and exertion of a usage right, multiple consumption actions (e.g. playing several songs) and obligation actions (e.g. deleting data) can happen, during this process. The attributes of the object (e.g. the amount of the 'not used' objects) and subject (e.g. balance in her/his account) are constantly changing. The authors defined models for attribute updating actions. Temporal constraints on usage actions and attribute updating actions are also formulated \cite{Zhang-2005-p351-387}.

The researches of Docomo Euro Lab and ETH Zurich propose taxonomy for the usage control policy related factors, e.g. usage activities, context attributes \cite{Pretschner2008}, obligations \cite{Hilty-2008-p531-546} etc. Their discussion about the mechanisms for inspecting low-level usage activities \cite{Pretschner2006,HarvanPretschner09,DBLPPretschnerRSW09,Berthold-2007-p18-25} is based on a holisthic viewpoint that exams the state changes related to consumed assets. Policy breach is defined as the status leading to information transference into unallowed descriptors, i.e. process, file on the disk, socket, memory area, etc. 

The usage control model is based on re-thinking the security policy with 'due usage' requirements. It involves improving traditional access control models by enriching the granted rights to accommodate various consumption activities and obligations \cite{Hilty-2008-p531-546,Pretschner2006}. Conditions for accessing rights can be based on single-credential (e.g. role) or multiple credentials (e.g. attributes). Our work follow such principle and focus on the case of multiple credential-based 'due usage' control scenarios. We also address the policy interaction and ratification problem which has not been explicitly explored in classical policy models.

\subsection{Attribute-based access control}\label{accincsss}
Attribute-Based Access Control ('ABAC' for short) policy model \cite{Wang-2004-p45-55,Wang2009a,Zhang2005,Cirio2007}. Rules in an ABAC policy are built through combination of attribute predicates with logical operators.
One of the representative work is XACML \cite{OASIS} standard, which defines a general purpose attributed-based access control language using the XML syntax. Access decision is based on the status (described by attributes) of object, subject, context and actions, as well as other information in the request context extracted by XQuery \cite{W3Cxq2011}. XML-based syntax makes it easy to be coordinated with other XML technologies, (e.g. it uses functions standardized in XQuery and XPath \cite{W3Ca} to build the attribute predicates), or processed with XML-based tools, e.g. DOM, JDOM, SAC, etc. Attribute-Based Access Control ('ABAC' for short) policy model bases the access condition on multiple credentials therefore is pertinent to network-based open system paradigms. Our work is align with the ABAC policy model in that the multiple-credential condition approach in our policy model is inherited from ABAC. But our policy model involves the method to resolve the issue of aggregating policies from different security domains, using ratification approach to avoid conflictions. To this end, we give a brief survey on policy weaving and ratification research in the following.

%The usage control model is based on re-thinking the security policy with 'due usage' requirements, to enforce security regulations aware of both the runtime environment and the assets usage. It involves making traditional access control models evolve:
%\begin{itemize}
%\setlength{\itemsep}{1pt}
%\setlength{\parskip}{0pt}
%\setlength{\parsep}{0pt}
%\item First, the granted rights are enriched to accommodate various consumption activities and obligations \cite{Hilty-2008-p531-546,Pretschner2006}.
%\item Second, access condition is multiple attributes-based, instead of traditional single-attribute (identity, role) based access control, therefore more capable to express various security attributes of system entities \cite{Zhang2008a,Park2004}
%\end{itemize}
%

%
%With the fundamental issues being handled from defining rights and entities attributes to ensuring policy correctness, usage control is developing toward a comprehensive model for expressing rich security profiles and controlling digital asset usage. Nevertheless, an important phenomenon emerges in today's cooperative applications that assets propagate and disseminate, crossing security domains, through aggregations and derivations. Therefore, providing full lifecycle protection with access control mechanisms involves the 'originator control' thoughts \cite{Park:2002:OCU:863632.883494,Ni2009a}, which means that the downstream consumptions activities on information should always get approval from the originators \cite{LGF2010rohtua}.

\subsection{Policy weaving \& lifecycle management systems}\label{cac}
A wide variety of asset access control systems for regulating assets propagation in cooperative and decentralized context have been proposed recently. A common feature of these systems is that the mechanisms for applying providers' policies upon the full lifecycle of assets must consider the impact of complex cooperative context, including asset aggregation, propagation and dissemination.
For example, much research attention has been paid on the control of information dissemination with policies \cite{LGF2010rohtua,Singh2008,DBLPSinghVBM08,Ni2009a,Li2006,Bandhakavi-2006-p51-58}. The dissemination paths usually form complex patterns as \emph{chain}, \emph{tree} \cite{Li2006} or \emph{graph} \cite{zsfbAPMS11}. Therefore approaches for dissemination control are closely related to the features of application context, allowing users to express the path their data can follow \cite{Bhargavan2008} and the usage restriction upon the data \cite{LGF2010rohtua}. In some circumstances (e.g. restricting event replay in publish/subscribe systems \cite{Singh2008,DBLPSinghVBM08}), filtration and distortion are needed to restrict the access to some information. It can be seen that the central issue of dissemination control is to cope with assets derivation and maintain the effects of originators' policies.

\subsection{Policy ratification}\label{plc_ratification}
A fundamental problem for a system with multiple policies is to find out whether a policy system meets the intention of policy authors (usually called policy \emph{ratification} \cite{AgrawalGLL2005} or policy \emph{safety} analysis \cite{Lindan2010}). It mitigates the policy 'leakage', where rights are granted to subjects that should be unauthorized, or 'over approximation', where subjects expected to obtain a right can actually never be authorized by the policy set.
The analysis tasks usually consist in:
\begin{itemize}
\setlength{\itemsep}{1pt}
\setlength{\parskip}{0pt}
\setlength{\parsep}{0pt}
\item \emph{Coverage} (or  \emph{totality} \cite{Bertolissi-2008-p217-225}), which is whether the policy covers the 'interest cases' (the conventional criteria is all the possible access requests in the system) \cite{Karat:2009:PFS:1850636.1850640} considered by the policy author. Example of this task is the 'policy effect query' in the EXAM policy analysis system \cite{Lindan2010}.
\item  \emph{Conflicts}, which means that some requirements for a right can not be achieved simultaneously \cite{Moffett93,Lindan2010}. If these requirements are all deemed obligatory, conflicts among them lead to inconsistency \cite{Bertolissi-2008-p217-225}.
\item  \emph{Dominance}, which refers to the circumstance where a rule adding to the system does not affect the system behavior, due to that other rules override (dominance) this rule. Then it's usually called a \emph{ineffective} \cite{AgrawalGLL2005} or \emph{redundant} \cite{PistoiaIBM2007} rule.
\end{itemize}

We discuss the method for detecting the relations of conflicts and dominance, based on the argument that they occur only when there is \textbf{policy overlapping}, i.e. several policies are defined on the same, or overlapped, set of entities. In our policy model an entity is defined on a set of attributes. Therefore, overlapping detection is based on comparing the attributes sets (of the same 'entity' category, i.e. subject, object, context, right, obligation or restriction) between policies.
Coverage analysis depends on not only the policy model, but also the capability of vocabulary base for describing the application domain knowledge (our vocabulary base organization will be discussed in future work).

The co-effect of overlapping rules is decided by the effect of each individual rule and the way they are combined, e.g. using \emph{combinators} to set precedence among multiple rules. Methods for combining rules can be differentiated along two dimensions:
\begin{itemize}
\setlength{\itemsep}{1pt}
\setlength{\parskip}{0pt}
\setlength{\parsep}{0pt}
\item \emph{Specificity precedence} \cite{Reeder2011}, which means that a rule applying to a more specific entity set takes precedence.
For example the file access policies in OS as Windows, Linux, etc. In such systems, if a policy $P_1$ defines that a group of files are allowed to be read by a user $U$, but one file in the group is defined by another policy $P_2$ as not readable by $U$. Then $P_2$ takes precedence.
\item \emph{Effect precedence} \cite{Li-2009-p135-144} decides the precedence among rules according to their effects -- deny or permit. Examples include:
\begin{itemize}
\setlength{\itemsep}{1pt}
\setlength{\parskip}{0pt}
\setlength{\parsep}{0pt}
\item \emph{One effect takes precedence}, for example \emph{deny override} (deny rule takes precedence) and \emph{permit override} in XACML \cite{OASIS}.
\item \emph{Majority takes precedence}, referring to that the effect (deny or permit) receiving more 'vote' (more rules with that effect) wins \cite{Li-2009-p135-144}.
\item \emph{Precedence by order}, as the \emph{first applicable} in XACML \cite{OASIS}.
\item \emph{Consensus}, which defines that all the rules should have the same effect to get a decision \cite{RLBLL2011,Li-2009-p135-144}. For example \cite{Li-2009-p135-144} proposed \emph{weak-consensus} and \emph{strong-consensus} strategies; the integration algebra in \cite{RLBLL2011} is based on the strong-consensus combining method.
\end{itemize}
\end{itemize}

Our aggregation algorithm follows the \emph{specificity precedence} principle and uses \emph{deny override} combinator to aggregate policies from different providers, based on the observation that only when approvals from all the providers of an aggregated asset are obtained that a right can be granted.
A recommendation method based on \emph{majority precedence} principle is proposed to facilitate the selection of more suitable partners from those having positive policy ratification results.

\section{Attribute-based usage control for cooperative computing context}\label{Toward}
An attribute-based usage control for cooperative computing context will accommodate rights upon resource and affiliated obligation in its 'authorization' component and attributes concerning the subject, object and context in its 'condition' component. This ensures a fine-grained control of resource usage and corresponding condition. Furthermore, to maintain upper-stream providers' security policy during assets aggregation, policy lifecycles management and policy aggregation is required for the ratification process of policies from multiple domains. Therefore, we can summarize the attribute-based usage control scheme as following.

\begin{definition}[attribute-based usage control Scheme]
an attribute-based usage control system is represented with a tuple:
\begin{equation}
ABUC= (O, Sh, Rt, S, Ct, Ob, Rn, P, V, Lc, G)
\label{scheme}
\end{equation}
where
\begin{itemize}
\setlength{\itemsep}{1pt}
\setlength{\parskip}{0pt}
\setlength{\parsep}{0pt}
\item \(O\) (Objects) represents the asset(s) protected by policy.
\item \(Sh\) (Stakeholder) represents the provider of an asset and associated policy.
\item \(Rt\) (Rights) includes the privileges upon the asset defined by $Sh$.
\item \(S\) (Subjects) represents the parties that can get rights on the assets.
\item \(Ct\) (Context) represents the collaboration context status associated to the system infrastructure, the environment and the business transaction.
\item \(Ob\) (Obligations) includes the achievement that must be done by the subject when it exerts the right.
\item \(Rn\) (Restrictions) represents the attributes (from $Ct$ set) associated to rights or obligations to further confine in what circumstance they are carried out.
\item \(P\) (Policy) is the policy definition. It builds deduction relations from a set of predicates on attributes of $O$, $S$ and $Ct$ to a set of predicates on $Rt$ or $Ob$.
\item \(V\) (Vocabulary) is a knowledge base collecting the attributes and their value ranges, according to application domain knowledge, that can be used by the policy model.
\item \(Lc\) (Lifecycle) is factors indicating the lifecycle of a policy, rule or predicate, to define whether their effects only cover partners directly interacting with the owner, or all downstream (indirect) consumers of the owner's assets.
\item \(G\) (Aggregation algebra) is the algebra used to combine the individual policies from each stakeholder, when their assets merge. The resulting policy reflects the co-effect of original policies.
\end{itemize}
\label{schemedef}
\end{definition}

an attribute-based usage control system takes into account the attributes of the assets, the consumers, the information system infrastructure and the collaboration context. It enables multiple providers to co-define the policy upon the artifact (C-Asset) of cooperative work. By tagging some (or all) rules in its policy with a 'lifecycle' ($Lc$) factor, a provider defines which rules (or the whole policy) should take effect beyond \emph{direct partner}, in other words, during the full lifecycle of an asset, even after it is merged with other assets. An algebra ensures that the policy aggregation is consistent and represents the co-effect of original policies.

%A 'design pattern' can be seen in recent policy models for open system contexts, the representative case being the P3P \cite{Cranor2002} project. In such approaches \cite{Cranor2002,LGF2010rohtua}, policies are used by information consumers (e.g. web sites) to express their 'promises' about how the providers' information will be used.

With this policy scheme, consumers use 'Quality of Protection' (\(QoP\)) policies to express their predefined promises about the protections that they offer to the assets. Therefore, $QoP$ policies include consumers' security profiles (represented by attributes) and the privileges they demand upon the assets. In this way, a request generated from the $QoP$ provides enriched information for the providers. We use 'Requirements of Protection' (\(RoP\)) policies to express the providers' requirements and evaluate the \(QoP\) with \(RoP\). Both \(QoP\) and \(RoP\) comply with the syntax we propose in the following.

\subsection {A concise syntax}\label{Basic}
We define an abstract syntax to present the attribute-based usage control policy model using EBNF (the version defined in ISO/IEC 14977).

\subsection {Basic syntax}\label{syntax}
An attribute-based usage control policy $P$ is a logical combination of rules $R$.
\begin{equation}
\begin{split}
P::=["\neg"],R,\{["\wedge"|"\vee"],["\neg"],R\}
\end{split}
\label{plcy}
\end{equation}

A rule is defined as:
\begin{equation}
R::=Rt, "\wedge "{,Ob},["\wedge "{,Rn}] ,"\gets ",C;
\label{rRUL}
\end{equation}
where:
\begin{itemize}
\setlength{\itemsep}{1pt}
\setlength{\parskip}{0pt}
\setlength{\parsep}{0pt}
\item \(Rt\) (Rights) denotes the operations upon the asset defined by the provider.
\item \(Ob\) (Obligations) defines the requirements (e.g. some operations) that the consumers must achieve if they exert the rights.
\item \(Rn\) (Restrictions) is attached to $Rt$ or $Ob$ to confine the run-time status related to their fulfillments.
\item \(C\) (Condition) summarizes the requirements that must be satisfied in order to grant rights to a consumer.
\end{itemize}

The $Rt$ element of a rule can involve multiple rights (formula \ref{R}). Negative symbol can be used, resulting in a negation rule.
%Rights can be refined with restriction (formula \ref{CT}), which gives more confining information.

\begin{equation}
\begin{split}
Rt::= "Rt(",["\neg"],[right], \{["\wedge"|"\vee"],["\neg"],right\},")";
\end{split}
\label{R}
\end{equation}

%\begin{equation}
%Rn::= "Rn(",["\neg"],[restriction], \{["\wedge"|"\vee"],["\neg"],restriction\},")";
%\label{CT}
%\end{equation}

\begin{equation}
Ob::= "Ob(",["\neg"],action,\{["\wedge"|"\vee"],["\neg"],action\}")";
\label{OB}
\end{equation}

%As an example, restriction $three\ times$ may be used to refine the right $rendering\ a\ piece\ of\ multi-media\ file$.
Rights and obligations can be refined with restriction, which gives more confining information.
An example of associating obligations to a granted right can be: if \emph{read\ client\ data} (a right) then \emph{delete acquired data} (an obligation) \emph{in 10 days} (a restriction).

Rights are released thanks to conditions related to either subject (consumer) attributes (denoted as \(SAT\)), object (asset) attributes (denoted as \(OAT\)) or context related attributes (denoted as \(CNAT\)), as shown in the following formula:
\begin{equation}
C::= ["\neg"],SAT,["\wedge"|"\vee"],OAT,["\wedge"|"\vee"],[CNAT];
\label{CN}
\end{equation}
where $SAT$, $OAT$ and $CNAT$ are the sets of predicates on attributes of subject, object and context. They are defined in the following formulae:
\begin{equation}
\begin{split}
SAT::=& "SAT(", ["\neg"], subject\_{}attribute\_{}name, operator, attribute\_{}value,\\
      & \{["\wedge"|"\vee"],["\neg"], subject\_{}attribute\_{}name, operator, attribute\_{}value\}, ")";
\end{split}
\label{SAT}
\end{equation}
\begin{equation}
\begin{split}
OAT::= & "OAT(", ["\neg"], object\_{}attribute\_{}name, operator , attribute\_{}value,\\
       & \{["\wedge"|"\vee"],["\neg"], object\_{}attribute\_{}name, operator , attribute\_{}value\}, ")";
\end{split}
\label{OAT}
\end{equation}
\begin{equation}
\begin{split}
CNAT& ::= "CNAT(", ["\neg"], context\_{}attribute\_{}name, operator , attribute\_{}value,\\
       & \{["\wedge"|"\vee"],["\neg"], context\_{}attribute\_{}name, operator , attribute\_{}value\}, ")";
\end{split}
\label{CNAT}
\end{equation}

Context attributes include the attributes of system infrastructure, environment and business session. An example is $organizationName=TechoInc.$,where the attribute name is $organizationName$, the operator is $equal to$ and the attribute value is $TechoInc.$.

\subsection {Rule lifecycle descriptors}\label{epsusecase}
Syntax elements $Sh$ and $Lc$ (see definition \ref{schemedef}) are related to policy aggregation process. $Sh$ indicates the owner of a policy, it provides provenance information after original policies are merged. The $Lc$ can be attached to a policy or to a specific attribute predicate in the policy to indicate the effect scope of the policy or predicate. In cooperative systems, policies can take effect beyond direct partners to protect assets providers' intellectual properties in the outcome of cooperative context. For example, by tagging a policy with $Lc=eot$ ($eot$ represents the end of transaction), the provider stipulates that the policy should be respected during the whole business process, by anyone that consumes an artifact containing the asset associated to that policy. Those predicates that only take effect between direct partners are attached with $Lc=dp$, where $dp$ stands for 'directly communicated partners'.

\section {The negotiation process}\label{Negotiation}
The negotiation process is described based on Abductive Constraint Logic Programming, obligations treated as condition, due to the observation that the fulfillment of obligations must be guaranteed before rights can be granted. The system entities' status changes induced by the grant or exertion of rights \cite{Craven-2009-p239-250} and obligations \cite{Cuppens12} can be described by Event Calculus.

\subsection{Request and Decision}\label{Request}
A request is generated from consumer's $QoP$. It declares the consumer's attribute (SAT), the attributes of the assets it demands (OAT), the context attributes (CNAT, which includes the attributes of the infrastructure, the environment and the business session), the rights it requires and the obligations it will align with. Our policy should consider state changes of system entities. Therefore the temporal information of the access-related actions, e.g. request generation, decision and response of policy evaluation, enforcement of decision, etc., should all be recorded. Such information is regulated as (events) attributes of the business session. They are regulated in the policy model, as discussed in section (\ref{State regulation}).
\begin{equation}
\begin{split}
req(&Rt,QoP_{AC}.SAT,QoP_{AC}.OAT,QoP_{AC}.CNAT,QoP_{AC}.Ob,t)
\end{split}
\label{rqst}
\end{equation}

When a request is matched by a rule in the policy, the effect of the rule is returned as a decision (usually the permit or deny of some rights). The decision should also indicate the subjects, objects, obligations and restrictions. The following formula illustrates this process.

\begin{equation}
\begin{split}
[\neg]Authorize(Rt,S,O,Ob,Rn,t) \gets QoP_{AC}\supset_{Sa}RoP_{AP}
\end{split}
\label{plruls}
\end{equation}

In above formula, $Rt$, $S$, $O$, $Ob$, $Rn$, $t$ stand for rights, subjects, objects, obligations, restrictions and 'time' respectively. It states that if the $QoP$ of asset consumer fulfills the $RoP$ of asset provider, rights can be granted, if obligations are imposed, and these rights can be followed by restrictions. Temporal factor $t$ indicates the time point when the negotiation decision is made and, therefore, will be used to exam the temporal restrictions. The function \(\supset_{Sa}\) denotes that the \(QoP_{AC}\) matches the \(RoP_{AP}\). This involves matching a request with each rule and then considering the effect of multi-rules related with a combinator.

\subsection {Matching mechanism}\label{mtching}
The foundation for two attributes to match is that they are certified by issuers trusted by each other, i.e. in one Certificate-Chain ($CC$). The formula (\ref{Tissuer}) describes this semantic.

\begin{equation}
\frac {\makebox[2.5in][l]{\begin{minipage}{4in}$\exists CC: issuer_{AR}\in CC,issuer_{AQ}\in CC$ \end{minipage}}}{issuer_{AR},issuer_{AQ}\models True}
\label{Tissuer}
\end{equation}

In this formula and the following, the syntactic element $AR$ stands for an attribute in the rules, $AQ$ for an attribute in the request. We use the notation convention that for an element $P$, $X_{P}$ refers to a child element (or property) of $P$. So $issuer_{AR}$ refers to the issuer of the $AR$.  $\models$ stands for the 'entailment' relationship.  $A \models_L X $ means that $X$ is provable from $A$ for a language $L$. In other words, $X$ is a semantic consequence of a set of statements $A$ under a deductive system that is \emph{complete} (all valid arguments are deducible -- provable) and \emph{sound} (no invalid arguments is provable) for a language $L$. The formula (\ref{Tissuer}) is necessary in a decentralized context (as a business federation) as the attributes of partners may be issued (certified) by different issuers. Trust among these issuers is necessary for the partners to acknowledge the status of each other.

The formula (\ref{aaMatch}) asserts that if the attribute names ($attr\text{-}id$) of $AR$ and $AQ$ are the same, they are in the same category (in other words, $SAT$ can only match with $SAT$, $OAT$ with $OAT$, and so on), their types (that is, data type of the attribute value) are the same and they are issued by trusted issuers, the $AR$ and $AQ$ can match each other, depending on the predicate function $fcn_{M}$ of $AR$ and the attribute value $AV$ provided by $AQ$, that is, if the attribute value provided by the request is in the attribute value range defined by the rule.

\begin{equation}
\frac {\makebox[4.0in][l]{\begin{minipage}{4.3in}$attr\text{-}id_{AR}=attr\text{-}id_{AQ}\ \wedge \
cat_{AR}=cat_{AQ}\ \wedge \
type_{AR}=type_{AQ}\\
\forall issuer_{AR},issuer_{AQ}: issuer_{AR},issuer_{AQ}\models True$ \end{minipage}}}{AR,AQ\models_{m}AV_{AQ}}
\label{aaMatch}
\end{equation}

The formula (\ref{pmatch}) states that for an attribute predicate $M$, if its attribute name $AR$ is matched by an attribute name $AQ$ in the request $Q$ and the $AV$ of $AQ$ satisfies $fcn_{M}(AV_{M})$ (that is, in the range of value defined by the predicate function $fcn_{M}$ and attribute value in $AR$), $M$ is matched by the $Q$.
\begin{equation}
\frac {\makebox[1.8in][l]{\begin{minipage}{2.1in}$\exists AR\in M: AR,Q\models_{m} AV_{AQ}\\ fcn_{M}(AV_{M},AV_{AQ})=True$ \end{minipage}}}{M,Q\models True}
\label{pmatch}
\end{equation}

The formula (\ref{rmatch}) describes the matching between a rule $R$ and a request $Q$: if all attribute predicates $M$s in $R$ are matched by $Q$, the access decision is provided by the effect of the rule, i.e. $Effect_{R}$.

\begin{equation}
\frac {\makebox[1.5in][l]{\begin{minipage}{2in}$\forall M_{R}: M_{R},Q\models True$ \end{minipage}}}{R,Q\models Effect_{R}}
\label{rmatch}
\end{equation}

\subsection {Obligation}\label{Obligation}
Obligations are deemed as constraints for future execution behavior of the system \cite{Dougherty-2007-p375-389}. The promise of obligation can be seen as a pre-condition for granting rights: \(grant(R)\gets fulfill (C)\wedge promise (Ob) \). The negotiation process must also evaluate potential conflicts between the obligations in request ($AQ$ in formula \ref{Obmatch}) with authorizations in the policy ($AR$ in formula \ref{Obmatch}):
\begin{equation}
\frac {\makebox[3.3in][l]{\begin{minipage}{4in}$\exists M_{Rt}\in Rt_{R}, AR\in M_{Rt}, AQ \in Ob_{Q}: \\AR,Ob_{Q}\models_{m} AV_{AQ}\ \ \ \ \ \  fcn_{M}(AV_{M},AV_{AQ})=False$ \end{minipage}}}{Rt_{R},Ob_{Q}\models False}
\label{Obmatch}
\end{equation}

\subsection{Rule combinator}\label{rulcomb}
\begin{figure}[htbp]
\centering
\includegraphics[width=0.5\textwidth]{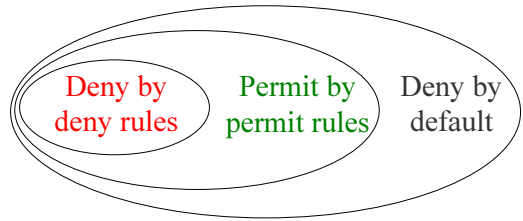}
\caption[requests partition]{Partition of requests domain with multi-layered effects} \label{fig:graph29}
\end{figure}

In situations where a request is matched by multiple rules, especially both permit rules and deny rules, a \emph{combinator} \cite{Ni-2009-p298-309,Li-2009-p135-144,Bruns-2007-p12-21,Bertolissi-2008-p217-225} is required to solve the conflicts.
We use the \emph{deny-override} combinator (in short, a request is always denied as soon as it matches a deny rule) \cite{OASIS} to get a stringent control strategy. By giving deny rules precedence, we get a more restrictive policy model, in order to avoid the unintentional rights granting due to that a slipshod permit rules takes precedent over the well-defined rules. Moreover, this strategy can be coupled with the 'negation as failure' principle to build a multi-layered effect space that allows fine-grained policy authoring (as illustrated in figure \ref{fig:graph29}).

The formula (\ref{eEff}) describes the evaluation of multiple rules coordinated by combinator. In the following formulae, the $\epsilon$ stands for the effect of rule ('deny' in our policy model), $P$ the policy combining several ($[1,n]$) rules $R$. A $P$ with $\epsilon\text{-}Overrides$ combinator is evaluated to effect $\epsilon$ if among all the rules $R$s that match $Q$, there are at least one rule $R$ with $\epsilon$ effect.

\begin{equation}
\frac {\makebox[2.1in][l]{$\exists i \in [1,n]: R_{i},Q\models \epsilon$}}{(P \ \epsilon\text{-}Overrides\   R_{1},...,R_{n}),Q\models \epsilon}
\label{eEff}
\end{equation}

The formula (\ref{IndeteEff}) asserts: The $P$ with $\epsilon\text{-}Overrides$ combinator is evaluated to $Indeterminate$ if no $R$ with $\epsilon$ effect has matched $Q$ and, at the same time, there is at least one $R$ that evaluates $Indeterminate$ with $Q$.
\begin{equation}
\frac {\makebox{\begin{minipage}{2.2in}$\forall R: R,Q\not\models Effect_{R}=\epsilon\\\exists j \in [1,n]: R_{j},Q\models Indeterminate$ \end{minipage}}}{(P \ \epsilon\text{-}Overrides\   R_{1},...,R_{n}),Q\models Indeterminate}
\label{IndeteEff}
\end{equation}

The formulae (\ref{neEff}) asserts: The $P$ with $\epsilon\text{-}Overrides$ combinator is evaluated to the opposite effect of $\epsilon$ (denoted as $\overline\epsilon$), if among all the $R$s that match $Q$, there is at least one rule $R$ with $\overline\epsilon$ effect, no $R$ with $\epsilon$ effect and, at the same time, all the $R$s that match $Q$ as $Indeterminate$ must have effect of $\overline\epsilon$.
\begin{equation}
\frac {\makebox[2.8in][l]{\begin{minipage}{4in}$\exists i: R_{i}, Q\models \overline\epsilon \\ \forall j: R_{j},Q\not\models \epsilon \\\forall R: R, Q\not\models Indeterminate\vee Effect_{R}=\overline\epsilon$ \end{minipage}}}{(P \ \epsilon\text{-}Overrides\   R_{1},...,R_{n}),Q\models \overline\epsilon}
\label{neEff}
\end{equation}

The formula (\ref{NAEff}) asserts: The $P$ with $\epsilon\text{-}Overrides$ combinator is evaluated to $NotApplicable$ if no $R$ matches $Q$.

\begin{equation}
\frac {\makebox[3.2in][l]{\begin{minipage}{4in}$\forall j: R_{j},Q\not\models \epsilon \\\forall j: R_{j},Q\not\models \overline\epsilon \\\forall j: R_{j},Q\not\models Indeterminate$ \end{minipage}}}{(P \ \epsilon\text{-}Overrides\   R_{1},...,R_{n}),Q\models NotApplicable}
\label{NAEff}
\end{equation}

According to the 'negation as failure' (close policy) principle in our model, the $Indeterminate$ and $NotApplicable$ evaluation results can be flattened to $Deny$ effect. But at implementation level, $Indeterminate$ results require usually compensation mechanisms, e.g. 'certificate discovery' processes to try finding the required information.

\subsection{Response}\label{Response}
A response can either deny or grant rights with some obligations and restrictions. In some situations, the asset attributes $OAT$ and the context attributes $CNAT$ can be modified.
\begin{equation}
[\neg]Authorize(Rt,S,OAT,CNAT,Ob,Rn,T)
\label{rspns}
\end{equation}

\subsection{State regulation}\label{State regulation}
When the consumer exerts the granted rights, the usage event (e.g. the assets consumption activities) usually leads to the system status changes (as the 'attribute updates' discussed in $UCON_{ABC}$ model). Such events should be considered in policy definitions, e.g. for the sake of 'model conflict detection'. We collect such events as 'business session attributes', which is part of the $CNAT$ ($CNAT$ consists in status information concerning the IT infrastructure, environment and business session, see section \ref{syntax}). Some attributes are dynamic (e.g. usage events, business session events or environment factors), whereas others (e.g. the infrastructure information) are relatively 'static' in the sense that they are not related to the usage or any other dynamic event that occurs in the business session context. The dynamic attributes in $CNAT$ can be described by event calculus \cite{Craven-2009-p239-250}.

For an incoming request, the context attributes set $CNAT$ includes the asset consumer's state information concerning its fulfillment of the rights (and associated obligations) granted to it before. As these context factors are related to the individual consumer, we denote them as \(CNAT_{ind}\).

There are also other 'public' contextual information that is not related only to one partner but rather to all the parters, concerning the state of the collaboration context as a whole. We denote this as \(CNAT_{pub}\).

Thus we have:
\begin{equation}
CNAT= CNAT_{ind}\cup CNAT_{pub}
\label{cnatind}
\end{equation}

The following formulae (\ref{permitted})-(\ref{violated}) define five types of \(CNAT_{ind}\) statements.
\begin{eqnarray}
permitted &(Rt,Sub,Ob,t)\ \ \ \ \ \ \ \ \ \ \ \ \ \ \ \ \ \ \label{permitted}\\
denied &(Rt,Sub,Ob,t)\ \ \ \ \ \ \ \ \ \ \ \ \ \ \ \ \ \ \label{denied}\\
obliged &(Act,Sub,Ob,t_s,t_e,t)\ \ \ \ \ \ \ \label{obliged}\\
fulfilled &(Act,Sub,Ob,t_{init},t_s,t_e,t)\label{fulfilled}\\
violated &(Act,Sub,Ob,t_{init},t)\ \ \ \ \ \ \ \ \label{violated}
\end{eqnarray}
where $Act$ is the set of actions and there is: \(Act \sqsubseteq Ob\), \(Act \sqsubseteq Rt\). $Sub$ is the subject, $Ob$ the object(s). Factor $t$ is the time point when the permission (formula \ref{permitted}), negation (formula \ref{denied}) or obligation (formula \ref{obliged}) is imposed. The $t_s$ and $t_e$ are the start time and the end time of the obligation respectively. The $t_{init}$ is the time point when the fulfillment or violation of the obligation happes. It can be seen that these \(CNAT_{ind}\) formulae express the system events that are regulated by policy, i.e. the events of authorizations (rights or obligations) and the fulfillment or violation of them.

Formulae (\ref{happen})-(\ref{broken}) define three types of \(CNAT_{pub}\) statements.
\begin{eqnarray}
happens &(event)\label{happen} \\
holdsAt &(event)\label{hodesat} \\
broken &(event)\label{broken}
\end{eqnarray}

They express events that are not managed by policy, e.g. events in business context as partner joining and quitting, session initiation, session termination, etc., or environmental events as changes of location, time, temperature, etc. Attribute-based usage control systems should be aware of state changes in the cooperative context caused by these events, as such changes lead to system entities' attribute updates.

Above semantics describes our policy evaluation process, which is aware of the context status. The following section describes a policy ratification process, which ensures that, when multiple policies are aggregated, the resulting policy represents the security goals of original policies.

\section{The ratification process}\label{UCONfC}
This section presents a ratification method (the semantics represented by an aggregation algebra), including conflicts detection and recommendation generation process, for policies from different security domains.

\subsection{An aggregation algebra}\label{Integration algebra}
As multi-policy ratification process is based on repeated peer-to-peer policy aggregation, therefore we first focus on the way that two policies are aggregated and use an aggregation algebra to formulate it.
\begin{definition}[Aggregation Algebra]
The aggregation algebra for cooperative policy is represented by a tuple:
\begin{equation}
(P, \Sigma, \Pi_{dc}, \&)
\label{IA}
\end{equation}
where $P$ is a set of policies, $\Sigma$ the vocabulary base consisting in the attribute names and value domains that the policies are defined on, $\Pi_{dc}$ a unary operation of domain projection, $\&$ a binary intersection operation.
\label{IntAlg}
\end{definition}

Our aggregation algebra is based on extensions of a fine-grained policy integration algebra \cite{RLBLL2011}:
\begin{itemize}
\setlength{\itemsep}{1pt}
\setlength{\parskip}{0pt}
\setlength{\parsep}{0pt}
\item The semantics of $\Sigma$ and $\Pi_{dc}$ are extended to multiple-owner policies aggregation.
\item $\&$ has been modified to apply 'strong consensus' to achieve 'originator control' \cite{Park:2002:OCU:863632.883494}.
\end{itemize}

Some definitions describe the basic concepts used in our policy scheme:
\begin{definition}[Access request]
Let $a_{1},a_{2},...a_{k}$ be a set of attribute names and $dom(a_{i})(1\le i \le k)$ the domain of possible value of $a_i$, let $v_{i}\in dom(a_{i})$. $Q\equiv\{(a_1,v_1),(a_2,v_2),...,(a_k,v_k)\}$ is a request over $\Sigma$. The set of all requests over $\Sigma$ is denoted as $Q_{\Sigma}$.
\label{Rq}
\end{definition}

\begin{definition}[3-valued access control model]
A 3-valued access control policy $P$ is a function mapping each request $Q$ to a value in $\{Y, N, NA\}$. The $Y$, $N$ and $NA$ stand for $Permit$, $Deny$ and $Not Applicable$ decisions respectively. $Q_{Y}^{P}$, $Q_{N}^{P}$ and $Q_{NA}^{P}$ denote the set of permitted, denied and not applicable requests by the P-based evaluation. $Q_{\Sigma_{P}}=Q_{Y}^{P} \cup Q_{N}^{P} \cup Q_{NA}^{P}$,\  $Q_{Y}^{P} \cup Q_{N}^{P}=\phi$,\ $Q_{Y}^{P} \cup Q_{NA}^{P}=\phi$,\ $Q_{N}^{P} \cup Q_{NA}^{P}=\phi$ (ensured by the 'deny-override' combinator in our basic policy model).
\label{IntAlg}
\end{definition}

Although access control system usually possess a 4-valued model \cite{Ni-2009-p298-309,OASIS} (i.e. including $Permit$, $Deny$, $Not Applicable$ and $Indeterminate$, more discussion on the effects of 4-valued policy model can be found in \cite{Ni-2009-p298-309}), we can deem $Indeterminate$ as $Not Applicable$ at policy model level discussion, as our policy scheme uses 'negation as failure' principle and any request that is not explicitly permitted is denied.

\begin{definition}[Domain constraint]
A domain constraint \textbf{dc} takes the form of \\ ${(a_{1},range_{1}),(a_{2},range_{2}),...,(a_{n},range_{n})}$, where $a_{1},a_{2},...,a_{n}$ are attribute names on $\Sigma$, and $range_{i}(1\le i \le n)$ is a set of values in $dom(a_{i})$. Given a request $r={(a_{r_{1}},v_{r_{1}}),...,(a_{r_{m}},v_{r_{m}})}$, we say that $r$ satisfies \emph{dc} if the following condition holds: for each $(a_{r_{j}},v_{r_{j}})\in r (1\le j \le m)$ there exists $(a_{i},range_{i})\in \emph{dc}$, such that $a_{r_{j}}=a_{i}$ and $v_{r_{j}}\in range_{i}$.
\label{dc}
\end{definition}

\textbf{\emph{Domain projection} ($\Pi_{dc}$)}: This operator takes the domain constraint \emph{dc} as a parameter to restrict a policy to the set of requests identified by \emph{dc} (see formula \ref{pidc}). Consequently, the domain projection operator $\Pi_{dc}$ delimits a set of attribute names and their corresponding value ranges. In a cooperative context such delimitation is necessary as partners from different organizations must share application domain vocabulary.
\begin{equation}
P_{I}=\Pi_{dc}(P)\Leftrightarrow \left\{
\begin{array}{l}
Q_{Y}^{P_{I}}=\{\mbox{$r|r\in Q_{Y}^{P}$\ and\ $r$\ satisfies\ $dc$}\}\\
Q_{N}^{P_{I}}=\{\mbox{$r|r\in Q_{N}^{P}$\ and\ $r$\ satisfies\ $dc$}\}
\end{array}
\right.
\label{pidc}
\end{equation}

The domain constraint means that when two policies are integrated, their vocabularies must comply with each other, precisely:

\begin{equation}
P_{I}=P_{1}\& P_{2}\Leftarrow (\Pi_{dc}(P_{1})\subseteq \Pi_{dc}(_{2}))\cup (\Pi_{dc}(P_{2})\subseteq \Pi_{dc}(P_{1}))
\label{pidc2}
\end{equation}

where \emph{\textbf{intersection}} ($\&$) is defined by formula (\ref{IAand}):

\begin{equation}
P_{I}=P_{1}\& P_{2}\Leftrightarrow \left\{
\begin{array}{l}
Q_{Y}^{P_{I}}=Q_{Y}^{P_{1}}\cap Q_{Y}^{P_{2}}\\
Q_{N}^{P_{I}}=Q_{N}^{P_{1}}\cup Q_{N}^{P_{2}}\\
Q_{NA}^{P_{I}}=Q_{Y}^{P_{1}}\cap Q_{NA}^{P_{2}} \cup Q_{NA}^{P_{1}}\cap Q_{Y}^{P_{2}}
\end{array}
\right.
\label{IAand}
\end{equation}

Given two policies $P_{1}$ and $P_{2}$, the intersection operation returns a policy $P_{I}$ which is applicable to all requests having the same decisions from $P_{1}$ and $P_{2}$. That is, $P_I$ permits (denoted as $Q_Y$) the requests that are permitted by $P_1$ and $P_2$ at the same time. It denies (denoted as $Q_N$) the requests that are denied by $P_1$ and $P_2$ at the same time. For all other cases, $P_I$ returns \emph{not applicable} (denoted as $Q_{NA}$) decision.

With the 'negation as failure' principle, we can further flatten the 3-valued algebra to the effects of $Permit$ and $Deny$, as all the $Not Applicable$ decisions result in the $Deny$ effect:
\begin{equation}
P_{I}=P_{1}\& P_{2}\Leftrightarrow \left\{
\begin{array}{l}
Q_{Y}^{P_{I}}=Q_{Y}^{P_{1}}\cap Q_{Y}^{P_{2}}\\
Q_{N}^{P_{I}}=Q_{N}^{P_{1}}\cup Q_{N}^{P_{2}} \cup Q_{NA}^{P_{1}}\cup Q_{NA}^{P_{2}}\\
\end{array}
\right.
\label{IAfn}
\end{equation}

This aggregation algebra is more explicit and reduces the complexity of aggregation algorithm design, as a request is permitted only when both providers permit it (i.e. when $Q_{Y}^{P_{1}}\cap Q_{Y}^{P_{2}}\not=\phi$). All other situations lead to the \emph{deny} decision.

This aggregation algebra is based on the request space and articulates our policy aggregation model. But during aggregation operations, directly comparing the request spaces of two policies is untractable, as enumerating all requests that can be accepted by a policy is not possible if some attributes predicates in the policy have unbounded value space (e.g. attributes defined on time, numeric ranges, etc.).

A request can be seen as a vector of attribute predicates. Therefore a natural switchover is to directly compare the attribute predicates. This allows to analyze the relations between the request spaces they cover, in order to get answers for questions like, e.g., "Does any request that is permitted by both of these policies exist?". Such thoughts lead to the study of relations between policy elements, namely 'rule similarity' analysis, which in turn is decided by the relations between attribute predicates in the rules.

\subsection{Relations between attribute predicates}\label{ap relation}
For two attribute predicates of the same category (i.e. subject, object, etc.), several relations can be identified.
\begin{itemize}
\setlength{\itemsep}{1pt}
\setlength{\parskip}{0pt}
\setlength{\parsep}{0pt}
\item \textbf{Contradict predicates}: predicates on the same attribute with disjunctive value domains (see formula \ref{contradict} for a natural deduction rule expression of it).
\begin{equation}
\frac {\makebox{\begin{minipage}{4.4in}$\forall AR_{1}\in M_{1}, AR_{2} \in M_{2}: cat_{AR_{1}}\models cat_{AR_{2}}\wedge attr\text{-}id_{AR_1}=attr\text{-}id_{AR_2} \\ \neg \exists AV_x: fcn_{M_1}(AV_{AR_{1}},AV_x)=True \wedge fcn_{M_2}(AV_{AR_{2}},AV_x)=True$ \end{minipage}}}{M_{1},M_{2}\models False}
\label{contradict}
\end{equation}

For example, two attribute predicates $lastAccess>10days$ and $lastAccess<3days$ are contradictable, as they define two value ranges $[10,\infty)$ and $[0,3]$ which have no intersection, upon the same attribute $lastAccess$. Therefore, when aggregating two positive rules $R_1$ and $R_2$, given that $Q_{Y}^{R_{1}}\not=\phi$ and $Q_{Y}^{R_{2}}\not=\phi$, we have $Q_{Y}^{R_{1}}\cap Q_{Y}^{R_{2}}=\phi$ only when $R_1$ and $R_2$ have contradictable attribute predicates.

\item \textbf{Distinct predicates}: attribute predicates that exist in only one rule (we omit the natural deduction rule expressions of this case and those following). %, e.g. the $encrypted=RSA$ in $RoP_C$ of the sample scenario (see formulae \ref{RoPC} and \ref{eRoPB}).

\item \textbf{Common predicates}: predicates existing in all the rules to be compared. %, e.g. the $certifier$ attribute of $SAT$ in the sample scenario (see formulae \ref{RoPC} and \ref{eRoPB}).

\item \textbf{Restricting predicates}: if two attribute predicates $M_1$ and $M_2$ (from different rules) have the same attribute name and the value range of $M_1$ is a subset of the value range of $M_2$, they form a pair of restricting predicates and $M_2$ is \textbf{\emph{restricted}} by $M_1$. For example $lastAccess<90days$ is restricted by $lastAccess<10days$.

%For example the $lastAccess$ in the $SAT$ elements in both $RoP_B$ and $RoP_C$ of the sample scenario (see formulae \ref{RoPC} and \ref{eRoPB}) belongs to such a case: $lastAccess<90days$ is restricted by $lastAccess<10days$.

\item \textbf{Intersecting predicates}: two attribute predicates (from different rules) that have the same attribute name, with different but intersected value ranges. An example of this case is: $lastAccess<10days$ and $lastAccess>3days$.
\end{itemize}

\subsection{Rule similarity}\label{Rule similarity}
According to the aggregation algebra, when aggregating two providers' policies, we focus on two problems:
\begin{itemize}
\setlength{\itemsep}{1pt}
\setlength{\parskip}{0pt}
\setlength{\parsep}{0pt}
\item finding out if there exists any request that can be permitted by both providers, according to the permit rules in their policies;
\item when such requests exist, finding out if they are all overridden by the deny rules of these two providers (in this case, no request can be permitted).
\end{itemize}

To find out the co-effect of rules, we define rule similarity according to a 'request space' viewpoint and summarize the following types of rule relation.

\begin{itemize}
\setlength{\itemsep}{1pt}
\setlength{\parskip}{0pt}
\setlength{\parsep}{0pt}
\item \textbf{Disjoint}:
If two rules have contradict predicates, their request spaces can never overlap. Therefore they are related by the disjoint relation, which means that no request can match both of them.

\item \textbf{Conjoint}:
If two rules have only common predicates, they cover the same request space. Therefore they are related by the conjoint relation. A request matching one of them matches also the other.

\item \textbf{Cover}:
For two rules $R_1$ and $R_2$, '$R_1$ covers $R_2$' means that all requests matching $R_2$ also match $R_1$. In other words, $R_1$ covers the request space of $R_2$. This occurs when $R_2$ restricts the value range of attributes in $R_1$, or when $R_2$ extends $R_1$ with new attributes. Consequently, we can identify three cases leading to the cover relation:
\begin{itemize}
\setlength{\itemsep}{1pt}
\setlength{\parskip}{0pt}
\setlength{\parsep}{0pt}
\item $R_2$ \textbf{restricts} $R_1$: each predicate in $R_1$ is restricted by a counterpart in $R_2$.
\item $R_2$ \textbf{refines} $R_1$: each predicate in $R_1$ has a counterpart in $R_2$ (with 'common predicate' relation) and, in addition, $R_2$ has some distinct predicates.
\item $R_2$ may both \textbf{restrict} and \textbf{refine} $R_1$ at the same time.
\end{itemize}

\item \textbf{Overlap}:
All other situations lead to overlap relation between the request spaces of two rules, which includes:
\begin{itemize}
\setlength{\itemsep}{1pt}
\setlength{\parskip}{0pt}
\setlength{\parsep}{0pt}
\item When there are intersecting predicates between the two rules. Because that in such situation none of them can cover the request space of the other defined on the attribute value domain of the intersecting predicates.
\item When both rules have distinct predicates (similar to above case, as none of them can cover the request space of the other associated to the distinct predicates).
\item When some predicates in a rule $R_1$ restrict predicates in rule $R_2$ and, at the same time, some predicates in $R_2$ restrict predicates in $R_1$.
\item The combination of the former 3 cases (all of the 3 cases can co-exist between two rules).
\end{itemize}
\end{itemize}

Rule similarity relations have direct impact on the aggregation process of policies that have multiple rules. For example, if there are 'disjoint' relations between two permit rules, the aggregation result can be matched by no request. Besides, a permit rule is 'covered' by a deny rule, the aggregation result of them will permit no request. The following section gives detailed discussion about aggregation.

\subsection{Aggregation algorithm}\label{Aggregation algorithm}
The aggregation process produces a 'Context Security Policy' ($CSP$), which consists in two sub-policy sets, \(RoP_{CSP}\) and \(QoP_{CSP}\). The \(RoP_{CSP}\) (see formula \ref{combineRoPcsp}) represents the requirements of the providers ($AP$ in formula \ref{combineRoPcsp}) whose assets are composed into the artifact of cooperative business process. It's defined as a combination of \(RoP'_{AP}\) which represents their policy elements (policy, assertion or attribute predicates) that are refined with the $Lc$ indicator. The $QoP_{CSP}$ combines the \(QoP_{P}\) (see formula \ref{combineQoPcsp}) of all the participants. It represents the context's guarantee about future participants' policies.
\begin{equation}
RoP_{CSP}=\sum_{i=1}^{m}(RoP_{CSP} \uplus RoP'_{APi})
\label{combineRoPcsp}
\end{equation}
\begin{equation}
QoP_{CSP}=\sum_{i=1}^{n}(QoP_{CSP} \uplus QoP_{Pi})
\label{combineQoPcsp}
\end{equation}

In formulae (\ref{combineRoPcsp}) and (\ref{combineQoPcsp}):
\begin{itemize}
\setlength{\itemsep}{1pt}
\setlength{\parskip}{0pt}
\setlength{\parsep}{0pt}
\item Factor \(m\) is the total number of asset providers in the collaboration context.
\item Factor \(n\) is the total number of partners (providers and consumers) in the collaboration context.
\item Operator $\uplus$ represents the aggregation function that uses the 'aggregation algebra' to aggregate two sets of policies.
\item Aggregated policies \(RoP_{CSP}\) and \(QoP_{CSP}\) are empty sets at the beginning.
\end{itemize}

We discuss the aggregation function in the following, paying attention to 'conflict' detection.

\subsubsection{Aggregation of permit rules}\label{Intperrul}
In order to integrate two permit rules, we have to pay attention to the rights predicates. Two permit rules can only be aggregated when they have overlapping 'rights' element, i.e. there are some rights which exist in both rules. If two permit rules, from different asset providers, define totally different rights, we call them \emph{\textbf{irrelevant permit rules}}. In such case, policy aggregation fails. Because that with 'negation as failure' and 'originator control' principles, a right upon a C-Asset is permitted only when all the providers permit it.

When the permit rules have overlapping 'rights' element, we have to check whether they are 'disjoint' rules (i.e. whether there are 'contradict predicates' between them, as 'contradict predicates' mean that no request can match both of rules).

When two rules are 'irrelevant' or 'disjoint', there is a \textbf{conflict of incompatible permit rules}. Otherwise, the two permit rules are \emph{\textbf{compatible}} and can be aggregated. In this case, all 'distinct predicates' and 'common predicates' between them are added to the CSP policy.  For a pair of 'intersecting predicates' or 'restricting predicates', the new value range is computed as the value range intersection of the two predicates in the pair. % (see predicate 'lastAccess=10days' in formula \ref{RoPCSP4} for an example).

%With this method, the aggregation result of $RoP_B$ and $RoP_C$ in the sample scenario (in section \ref{Towards}) is:
%
%\begin{equation}
%\begin{split}
%RoP_{CSP4}&.(Lc=eot):\\
%&Rt(actionID=read\vee actionID==merge)\\
%&\wedge Ob(actionID=delete\wedge actionTarget=O\\
%&\ \ \ \ \ \ \ \wedge actionTime=30days)\\
%&\gets\\
%&\ \ \ \ Sh(C=30,B=70)\\
%&\ \ \ \ \ \ \ OAT(ID=ME(A)\wedge encrypted=RSA)\\
%&\ \ \ \ \wedge SAT(certifier=A \wedge lastAccess=10days)\\
%&\ \ \ \ \wedge CNAT(deliveryChannel=SSL).(Lc=dp(4))
%\end{split}
%\label{RoPCSP4}
%\end{equation}
%
%In formula (\ref{RoPCSP4}), $Lc=dp(4)$ indicates that the predicate $delivery\_{}channel=SSL$ is added in at 'the aggregation step 4'. It will be discarded after next aggregation step (an 'aggregation step' is associated to one information exchange step between partners in a business process).

This procedure describes the basic aggregation process for two policies both having only \textbf{\emph{one}} permit rule. When aggregating two policies both having \emph{\textbf{multiple}} permit rules, the aggregation process will form a series of permit rule pairs by selecting one rule from each policy. Then for each rules pair, it checks whether there is a conflict of incompatible permit rules. If all the rules pairs have such a conflict, the aggregation fails. Otherwise, compatible rule pairs are integrated to the resulting CSP policy.

\subsubsection{Aggregating policies with deny rules}\label{multi-permit}
When there are deny rules in the policies that are aggregated, the potential of conflict increases. For example if all permit rules are 'covered' (or 'overridden' as defined in XACML) by deny rules, the resulting policy can not authorize any access request. Generally, there are 2 types of conflict brought by deny rules in our policy models:

\begin{itemize}
\setlength{\itemsep}{1pt}
\setlength{\parskip}{0pt}
\setlength{\parsep}{0pt}
\item \textbf{Positive-Negative Conflict of Modalities}: It occurs when a subject is both authorized and prohibited for a right on an object. As we use deny-override combinator (see section \ref{rulcomb}), a permit rule covered by a deny rule is \textbf{\emph{ineffective}} (its effect is overridden by the deny rule). Therefore when aggregating two policies, after processing all the 'compatible permit rule' pairs, we should check if all the resulting rules generated from aggregating compatible rules are covered by some deny rules. In such case, the resulting CSP policy will not authorize any access request and the aggregation fails.
\item \textbf{Conflict between Imperial and Authority Policies}: It occurs when a subject is required to carry out an action by the 'obligation' of a policy and is prohibited to carry out this action by another policy. If this happens during the aggregation of policies from different partners, it means that the partners have contradictable security policies. Consequently they should not be gathered in one collaboration context. When the conflict occurs between policies belonging to a single owner, it denotes the inconsistent policy authoring work. As a consequence, its correction is left to the policy owner system, e.g. relying on policy ratification  \cite{AgrawalGLL2005} or policy safety analysis \cite{Lindan2010} systems.
\end{itemize}

\subsubsection{Other conflicts}\label{cnflcts}
Some other types of conflict in a cooperative context should be handled, not by the aggregation process itself, but by other methods. We give a brief discussion of them:
\begin{itemize}
\setlength{\itemsep}{1pt}
\setlength{\parskip}{0pt}
\setlength{\parsep}{0pt}
\item \textbf{Conflict of Duties}: It indicates the cases where two actions are not allowed to be performed by the same subject, e.g. by the 'separation of duties' principle \cite{Li2007}. Usually the policy author applies this principle by introducing a deny rule to its policy set to forbid some actions based on the past actions of the subject. This historical action information is deemed as attributes of the subject, in our policy model, and is provided by session events (including usage activities) collected by the policy enforcement system.
\item \textbf{Conflict of Interests}: It means that a subject is not allowed to perform the same action upon two objects. This is handled with the same approach as the 'conflict of duties'. The owner of the object has just to define a deny rule to express this principle.
\item \textbf{Conflict of Priorities for Resources}: It happens when more than one exclusive usage requests are made to a resource, or when the resource is associated to a limited amount, which is exceeded by the total amount demanded by all the requests. The 'pre-update' strategy of attribute management in $UCON_{ABC}$ \cite{Park2004} can ensure that such conflict is avoided. This strategy allows the attributes of subject and object (and any other related entity, if needed) to be updated by the access control system immediately after the right is granted and before it is exerted.
\end{itemize}

When more than one partner can be chosen to join the collaboration context and if none of them has a policy that conflicts with the CSP, a 'general majority voting' strategy like those in \cite{Ni-2009-p298-309} can be used to choose the most fitting partner w.r.t. security perspective, as introduced in the following.

\subsubsection{Aggregation recommendation}\label{Recmdtn}
From the aggregation algebra and the aggregation algorithm given previously, we can see that the CSP becomes more restrictive when aggregating new partners policies: rights upon the C-Asset decrease, whereas the condition expends. Consequently, the chance to find eligible future participants diminishes. We propose a partner recommendation method as a countermeasure. Basically, if more than one partner can be chosen to join the context ( fitting to the business goal and security requirements of the context), the partner with the more suitable security profile is chosen, according to a 'weighted majority voting' principle, based on extending the 'Majority Voting' \cite{Ni-2009-p298-309} method. 'Weighted' is due to the fact that we give more importance to $RoP$ of the partner than to the $QoP$, and for comparing $RoP$ policies, the importance among rule elements is weighted as: $condition>rights>restriction>obligation$.

Therefore, we exam policies according to the following priority:
\begin{itemize}
\setlength{\itemsep}{1pt}
\setlength{\parskip}{0pt}
\setlength{\parsep}{0pt}
\item{Slim $RoP$}, which favors the partners that introduce less restricts to the $CSP$, thus allowing more opportunities for new consumers in the future step. A 'Slimmer' $RoP$ is decided with the following priority among policy elements:
\begin{description}
\setlength{\itemsep}{1pt}
\setlength{\parskip}{0pt}
\setlength{\parsep}{0pt}
\item [1] Slim $condition$ element, which has \textbf{fewer} attribute predicates that have \textbf{wider} value ranges. With fewer predicates, a rule defines a less specific (wider) range of entities (entities can be subject, object, context, environment, etc.), allowing the rule to match more requests (e.g. it's easier to match a rule with 5 attribute predicates than to match a rule with 50 attribute predicates). A wider value range of an attribute predicate covers a larger scope in the value domain. Thus it's easier for the request to fall in (match with) the range (e.g. $age \subset [1,15]$ covers a larger request space than $age \subset [1,3]$).
    The condition element has three components: subject, object and context. Each of them can have several attribute predicates, which can be further weighted. But the weight on a specific set of attributes is closely related to the domain of application and also depends on the policy author's goal. Therefore it should be left to be customized by users of the policy implementation system.
\item [2] Rich $rights$ element, which has more rights or the rights that give 'more usage opportunity' (closely depending on the application domain).
\item [3] Slim $restrictions$ element, similar to the $condition$ element.
\item [4] Slim $obligations$ element, similar to the $condition$ element.
\end{description}
\item{Rich $QoP$}, which favors the partners that make the aggregated $CSP$ have slim $rights$ element and rich $condition$, $restrictions$ and $obligations$ elements. It increases the chance to fit to $RoP$ policies for the future steps.
\end {itemize}

The rational of these principles is to "slow down" the growth of the \(RoP_{CSP}\) and the reduction of \(QoP_{CSP}\). The principles we've defined steer the combination of policies in a single step. To optimize the aggregated $CSP$ on the global scale of cooperative context (which involves multiple steps), the method presented in \cite{Lindan2010} can be used. In this system a policy is represented by a Multi-Terminal Binary Decision Diagram (MTBDD). Combining multiple policies means combining MTBBDs to form a Combined MTBDD (CMTBDD). As each CMTBDD represents an aggregation in our policy scheme, aggregation recommendation can be done by using state-of-the-art optimization methods to choose the best CMTBDD with our recommendation principle (slim $RoP$ and rich $QoP$).

%\subsection{Security management through CSP}\label{CSP evolve}
%The \(CSP\) evolves as participants join and quit. When new providers are required to join, there will be a policy negotiation between the \(QoP_{CSP}\) and the new provider's \(RoP\). Whereas before a consumer joining, the negotiation focuses on its \(QoP_{AC}\) and the \(RoP_{CSP}\). When a participant wants to quit, it has to fulfill any obligation imposed by the \(CSP\). The \(CSP\) can also change if a participant uses a 'release' operation to abolish its \(RoP\) upon the C-Asset.

\section{Conclusion}\label{cnln}
Attribute-based usage control policy model provides fine-grained control of resource access condition and continuous regulation of partners' behavior. With policy ratification and aggregation mechanism, it forms a foundation for security configuration of cooperative computing context. The paper describes an attribute-based usage control policy scheme, providing a concise syntax with EBNF and discussed the negotiation and aggregation processes.

The negotiation process is described based on Abductive Constraint Logic Programming, disambiguating the request-policy fulfillment relation by elaborating the attribute matching mechanism and effects of rule combinator.

The policy ratification process supports \emph{specificity precedence} principle and \emph{deny override} combinator to ensure that the resulting policy represents the co-effect of original policies. Basically, an access request to an C-Asset (artifact of the cooperative business process) is refused if any provider of resources (denoted as O-Assets) involved in the C-Asset refuses the request. An aggregation algebra is developed to describe the semantics of aggregation process, based on modification and extension of a previous policy integration algebra  \cite{RLBLL2011}. We give further discussion of attribute predicates relations and rule similarities, which have direct impacts on the policy aggregation method. A \emph{majority voting} strategy is proposed to recommend partners with preferable security profiles.

%In sum, our policy model provides the foundation for a 'peer to peer' style security governance in cooperative context. Thanks to features adopted from the usage control model, it can manage the security profiles of the asset, consumer, IS infrastructure, environment and business context, paying particular attention to asset usage actions and obligation fulfillment, as well as coping with assets aggregation/dissemination among partners in a cooperative business process.

Future work consists in the development of security governance framework based on the policy scheme, taking into consideration the impacts of complex cooperative contexts and particular IT infrastructures.

%\vfill
\bibliographystyle{ieeetr}
{\small
\bibliography{suziyi}

\end{document}
This is never printed